\documentclass[twocolumn,showpacs,superscriptaddress,aps,prl]{revtex4}
\usepackage[english]{babel}
\usepackage{xspace}
\usepackage{times,amsmath,amsfonts,amsthm}
\usepackage{graphicx}
\usepackage{color}
\usepackage{epsfig}
\usepackage{wrapfig}
\usepackage{dcolumn}
\usepackage{bm}

\newcommand{\beq}{\begin{equation}}
\newcommand{\eeq}{\end{equation}}
\definecolor{red}{rgb}{1.0,0.0,0.0}

\begin{document}
\title{Equilibrium Shape and Size of Supported Heteroepitaxial
Nanoislands}

\author{J. Jalkanen}
\affiliation{Department of Engineering Physics, P.O. Box 1100, Helsinki
University of Technology, FIN-02015 TKK, Espoo, Finland}

\author{O. Trushin}
\affiliation{Institute of Physics and Technology, Yaroslavl Branch,
Academy of Sciences of Russia, Yaroslavl 150007, Russia}

\author{E. Granato}
\affiliation{Laborat\'orio Associado de Sensores e Materiais,
Instituto Nacional de Pesquisas Espaciais, 12245-970 S\~ao Jos\'e
dos Campos, SP Brasil}
\affiliation{Department of Physics, P.O. Box 1843, Brown
University, Providence, RI 02912-1843}

\author{S. C. Ying}
\affiliation{Department of Physics, P.O. Box 1843, Brown University, Providence, RI 02912-1843}

\author{T. Ala-Nissila}
\affiliation{Department of Engineering Physics, P.O. Box 1100, Helsinki
University of Technology, FIN-02015 TKK, Espoo, Finland}
\affiliation{Department of Physics, P.O. Box 1843, Brown
University, Providence, RI 02912-1843}

\date{September 2, 2008}

\begin{abstract}

We study the equilibrium shape, shape transitions and optimal size
of strained heteroepitaxial nanoislands with a two-dimensional
atomistic model using simply adjustable interatomic pair potentials.
We map out the global phase diagram as a function of
substrate-adsorbate misfit and interaction. This phase diagram
reveals all the phases corresponding to different well-known growth
modes. In particular, for large enough misfits and attractive
substrate there is a Stranski-Krastanow regime, where nano-sized
islands grow on top of wetting films. We analyze the various terms
contributing to the total island energy in detail, and show how the
competition between them leads to the optimal shape and size of the
islands. Finally, we also develop an analytic interpolation formula
for the various contributions to the total energy of strained
nanoislands.

\end{abstract}


\maketitle

\section{Introduction}

The shape and size of adatom islands resulting from growth
processes has been a subject of numerous recent experimental and
theoretical studies
\cite{politi00, daruka99, daruka02, Wang2000, zangwill93, uemura02, katsuno05, Johnson97, spencer01, tersoff84, Budiman, Chen1998, Rudin1999, Combe2001}.
In the case of Stranski-Krastanow (SK) growth mode with
islands growing on top of a wetting film, the relative abundances
of different crystal shapes in a distribution show clear peaks at
certain volumes, known as optimal sizes or magic numbers. In
particular, when the growing islands are of nanoscopic size the
central issue is the possible spontaneous self-organization of
islands into arrays with a narrow size distribution. Such cases
offer immediate technological applications in modern
nanotechnology.

In spite of numerous publications in this field published in recent
years some aspects of such self-assembly process are still not quite
clear. In particular, there are still uncertainties as to whether
the observed shapes and sizes of growing islands in heteroepitaxy
correspond to thermodynamic equilibrium state of minimum free energy
or limited by kinetic effects.

In this paper, we examine the minimum energy configuration of an
array of islands with the given constraint of a fixed number of
adsorbate atoms and with a fixed island density determined by
initial growth conditions \cite{Wang2000}. There are various
physical mechanisms leading to the optimal size and shape of the
islands. Among them are the relevant surface tensions, adsorbate substrate
bonding, elastic relaxation in the island, wetting film thinning
and nonlinear elastic contributions, such as bending, buckling and
dislocations etc. \cite{Parry06,Spencer00,spencer01} The wetting
film and substrate can also mediate elastic dipole interactions
between the islands, which can play an important role
\cite{Shchukin1995}.

Most of the previous theoretical investigations of this problem
employ a continuum approach to treat the elastic properties of the
adsorbate and substrate. In addition, a predefined set of facets is
usually assumed for the islands together with a somewhat arbitrary
separation of surface and bulk terms in total energy expression.

The accuracy of these approaches for nanosized islands is yet to be
determined \cite{Spencer03,Miller00}. To avoid any such
approximations, in this work we use a fully atomistic model allowing
for both elastic and plastic strain relaxation without assumptions
on predefined shapes. To find the ground state of the atomistic
systems we apply here the molecular static approach, which
corresponds to the zero temperature limit. We adopt a
two-dimensional (2D) model, but extension of many of the results
presented here to the more realistic 3D systems is also possible.
The reduced dimension allows to study all the possible
configurations within feasible computer time.

This paper is organized as follows. First, we focus on a few
specific island shapes found for various material parameters by
searching for the minimum energy configuration in the small size
limit. Then by focusing on these island shapes and by varying the
misfit and the substrate-adsorbate interactions, we find
configurations corresponding to the commonly known different modes
of adsorbate growth, Frank-Van der Merwe (FM), Stranski-Krastanow
(SK) and Volmer-Weber (VW). For the SK mode, we demonstrate the
existence of optimal size and shape for the islands and study how
they vary with the coverage of the adsorbate. We discuss the
physical mechanisms that give rise to the narrow distribution of the
optimal shape and size. Finally, we present an interpolation formula
for the total energy which yields an accurate description of the
phase diagram and optimal size and shape of the island in comparison
with the numerical data. Moreover, this formula allows a clear
interpretation of the various competing strain energy relaxation
mechanisms.

\section{Model and Methods}

In this work, we adopt a 2D classical atomistic model
\cite{jalkanen05,tru02a,tru02b}, which has been previously used to
study microscopic mechanisms for strain relaxation in thin films.
This model also allows for an straightforward extension to the more
realistic 3D case \cite{Trushin06}.

Within the model atomistic system is relaxed through standard MD
cooling procedure and different island shapes are systematically
compared for finding minimal energy shapes. In 2D all relevant
island configurations can be studied within a reasonable computer
time. The model allows for both elastic and plastic strain
relaxation in heteroepitaxial system without any assumptions on
predefined shapes of the islands. Some of our results here are
obtained for relatively small, nanoscopic islands up to a few
hundred atoms in size, in order to examine deviations from the
continuum theory of elasticity \cite{Daruka1997,daruka99,daruka02}.

To allow easy adjustability and anharmonic effects, the interactions
between all atoms in the system are described by a modified
Lennard-Jones (LJ) pair potential \cite{zhen83} $V(r)$ with two
parameters, namely the dissociation energy $\varepsilon_{ab}$ and
the atomic equilibrium distance $r_{\rm ab},$
\begin{equation}
\label{LJpot}
V_{\rm ab}(r) = \varepsilon_{\rm ab}\left[\frac{5}{3}\left(\frac{r_{\rm ab}}{r}\right)^8 -
\frac{8}{3}\left(\frac{r_{\rm ab}}{r}\right)^5\right]\Theta_{r_{\rm ab},r_{\rm c}}(r),
\end{equation}
where
\begin{equation}
\Theta_{r_{\rm ab},r_{\rm c}}(r) = \left\{\begin{array}{cl}
1, &  r \leq r_{\rm ab};\\
3\left(\frac{r_{\rm c} - r}{r_{\rm c} - r_{\rm ab}}\right)^2 -
2\left(\frac{r_{\rm c} - r}{r_{\rm c} - r_{\rm ab}}\right)^3, & r_{\rm ab} \leq r \leq r_{\rm c};\\
0, & r_{\rm c} \leq r.
\end{array}\right.
\end{equation}
The modification ensures that the potential and its first derivative
vanish at the cutoff distance $r_{\rm c}$. The index $(\rm ab)$
stands for substrate--substrate $(\rm ss)$, substrate--film $(\rm
sf)$ or film-film $(\rm ff)$ respectively.

The parameter $r_{\rm fs}$ for the adsorbate-substrate interaction
is simply set as the average of the film and substrate lattice
constants, {\it i.e.} $r_{\rm fs}=(r_{\rm ff}+r_{\rm ss})/2$. The
lattice misfit $f$ between the adsorbate and the substrate can be
defined as
\begin{equation}
f=(r_{\rm ff}-r_{\rm ss})/ r_{\rm ss}.
\end{equation}
A positive mismatch $f>0$ corresponds to compressive strain and
negative $f<0$ to tensile strain when the adsorbate island is
coherent with the substrate.

As for the interaction parameters, the film--film and
substrate--substrate interactions are set to be equal, with
$\varepsilon_{\rm ff} =\varepsilon_{\rm ss}$.  The film substrate
interaction $\varepsilon_{\rm sf}$ can be parameterized by
 $\kappa$ defined as

\begin{equation}
\kappa =
{(\epsilon_{\rm ss} - \epsilon_{\rm sf})}/{\epsilon_{\rm ss}},
\end{equation}
\noindent
A negative value of $\kappa<0$ corresponds to an
effectively attractive and positive $\kappa>0$ to a repulsive
substrate.

Calculations were performed with periodic boundary conditions for
the substrate in the direction parallel to the
ad\-sor\-ba\-te\--sub\-stra\-te interface. The size of the unit cell
$l_w$ (measured in units of substrate lattice constant $r_{\rm ss}$)
determines the density of the islands in the SK and VW regime. Two
bottom layers of the substrate were held fixed to energy minimize a
se\-mi-in\-fi\-ni\-te substrate while all other layers were free to
relax. In most calculations the thickness of the substrate was $15$
layers, and additional tests were carried out to ensure that the
thickness did not influence the results.

\section{Results}

\subsection{Phase Diagrams in the Submonolayer Regime}

\begin{figure}[t]
\centering\epsfig{file=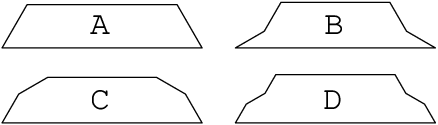}
\caption{\label{shapefig} The generic island shapes A, B, C and D as discovered for
nanoscopic islands \cite{jalkanen05}. All the shapes have at most a
single high energy facet.}
\end{figure}

The first issue of interest concerns the equilibrium shapes of small
islands in the submonolayer regime. In Ref. \cite{jalkanen05} we
presented a complete phase diagram as a function of the misfit $f$
and the total number of adsorbate atoms $N$ in this regime with
$\kappa = 0.$ It was found that there are four characteristic shapes
A, B, C, and D as shown in Fig. \ref{shapefig}. Of these the shape
D, which was not predicted by continuum theory calculations
\cite{Daruka1997,daruka98,daruka02}, was encountered only
occasionally for larger volumes and misfits. Here, we present the
corresponding results for the cases $\kappa = + 4 \%$ and $\kappa =
- 4 \%$, for $N\le85$ and the size of the periodic cell $l_w=200$.
According to our tests, the interaction parameter is the most important 
factor in determining the various growth modes in equilibrium.

Following the procedure explained in Ref. \cite{jalkanen05}, we find
the ground state shape of an island for a fixed total number $N_A$
of adsorbate atoms without assuming any predetermined shapes by a
systematic search approach. Each coherent configuration is described
by a set of integer numbers, ${n_i}$ specifying the number of atoms
in successive island layers. In terms of these numbers, the two
types of facets, considered in the previous works
\cite{daruka99,daruka02} correspond to $n_i - n_{i+1}=1$ for steep
facets and $n_i - n_{i+1}=3$ for shallow facets. The only physical
restrictions we impose are that the island has a reflection symmetry
about a line through the center and that overhangs are not allowed. Then,
for each initial configuration, molecular dynamics (MD) cooling is
run to allow the system to relax and reach a minimum energy
configuration. The equilibrium shape for a given $N_A$ is identified
as the relaxed island configuration with lowest energy among all the
configurations.

\begin{figure}[t]
\begin{tabular}{cc}
\centering\epsfig{file=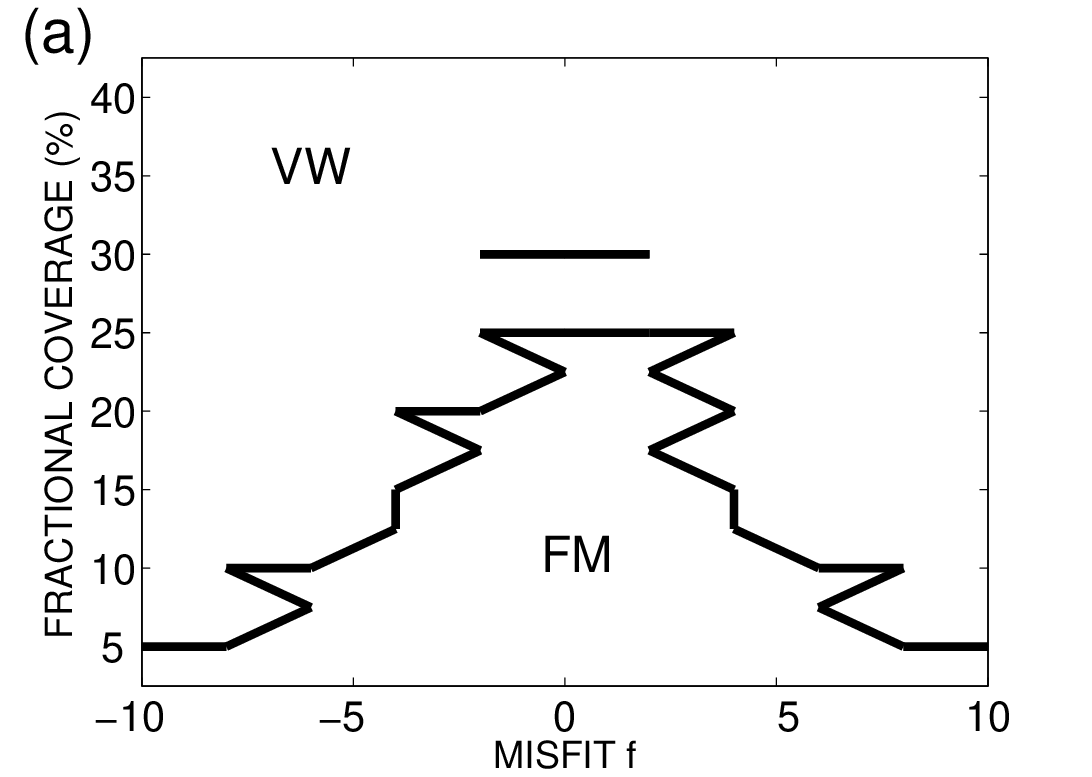,width=44mm} &
\centering\epsfig{file=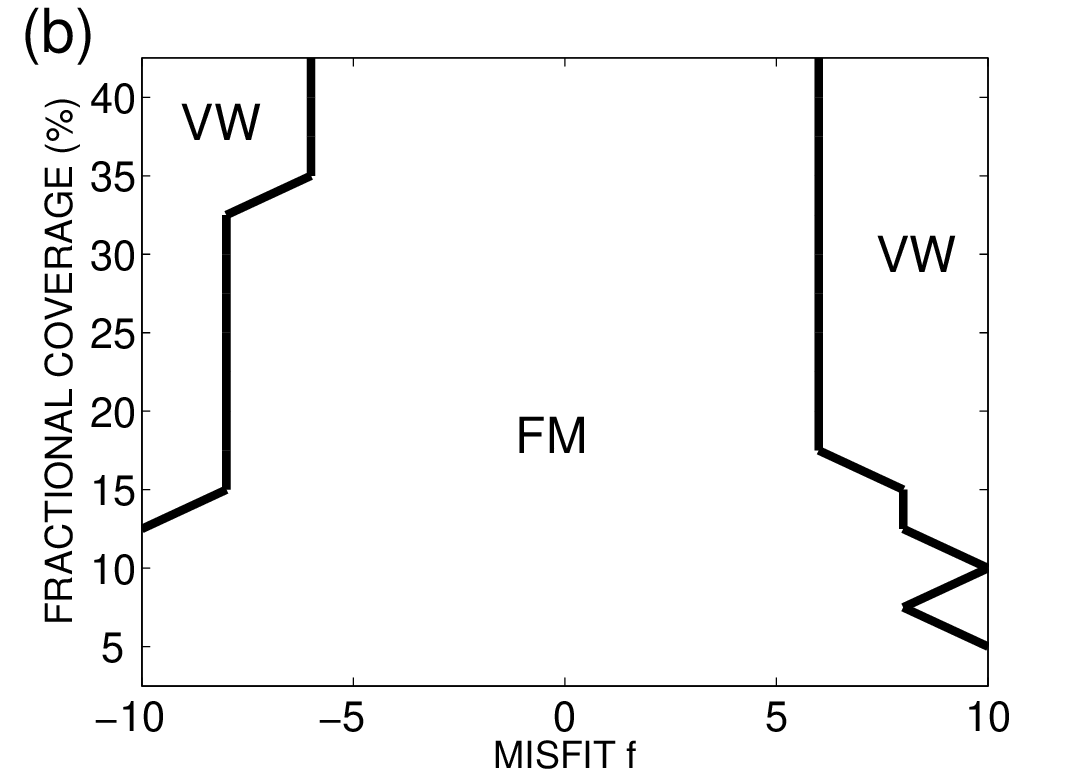,width=44mm}
\end{tabular}

\caption{Phase diagrams for submonolayer regime ($N_A \le 85$) with
$l_w=200$. Panel (a) is for $\kappa = + 4\%$ and (b) for the case
$\kappa = - 4\%$. The line in panel (a) is a small domain of
configurations that belong to the Frank\-- van der Merve regime (FM).
As expected, in the attractive substrate case island formation and
re\--entrant transitions are suppressed for fixed $f$ when $N_A$
increases. See text for details.\label{kappanonzero}}

\end{figure}

The  phase diagram for the case where the substrate is repulsive
with $\kappa= + 4 \%$ is shown in Fig. \ref{kappanonzero}(a). As
expected, island formation is enhanced as compared to the case where
$\kappa = 0.$ Even with zero misfit, island formation starts around
$N_A \approx 50$ and the wetting film disappears completely beyond
$N_A \approx 75$. We detect no new island shapes or any occurrences
of the shape D.

The results above should be contrasted to the case of an attractive
substrate with $\kappa = -4\%$ in Fig. \ref{kappanonzero}(b).
Complete wetting regime dominates the phase diagram almost up to the
largest values of mismatch considered here. As for the shape of the
islands, all but one case belong to the categories A and C. The
single exception at $f = -10\%$ and $N_A = 80$ falls in the category
D. We also note that as in Ref. \cite{jalkanen05}, the phase
diagrams here are asymmetric with respect to tensile and compressive
strain highlighting the importance of the anharmonicity of the
atomic interaction potentials.

\subsection{Global Phase Diagram at Higher Coverage }

At higher coverages with $N\gg l_w,$ it is possible to have in
addition to the partial wetting film (FM) and small island (VW)
phases also the phase where islands grow on top of wetting films
(SK).

\begin{figure}[t]
\centering\epsfig{file=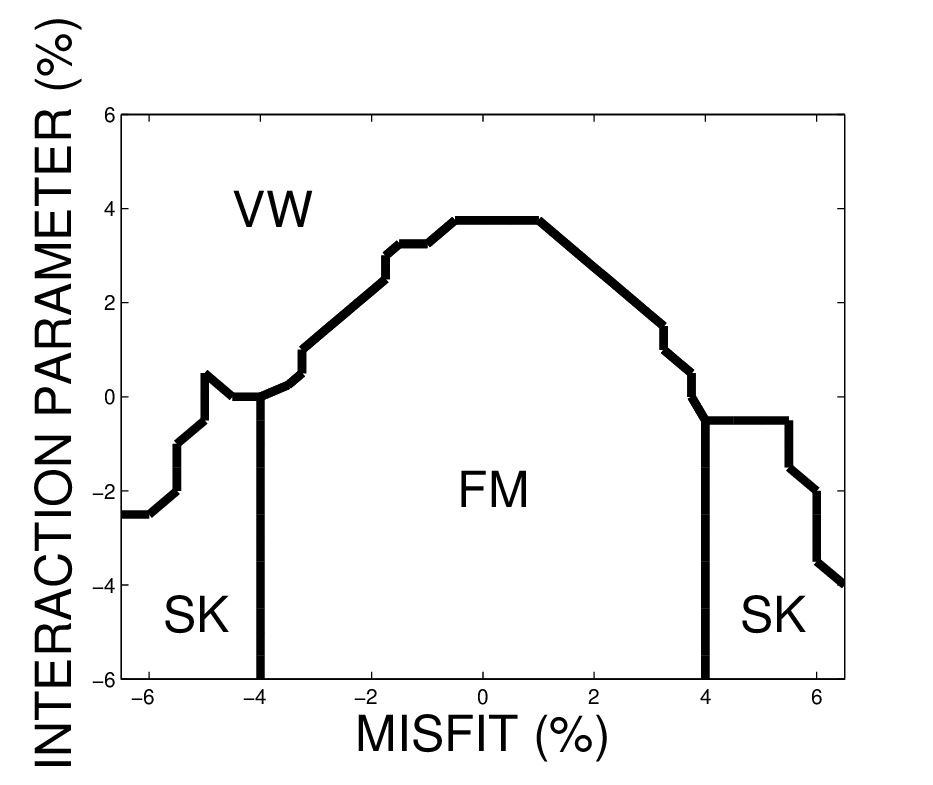,width=92mm}

\caption{\label{global_phase_diagram}
Global phase diagram of island
shapes as a function of the misfit parameter $f$ and the interaction
parameter $\kappa$ for $N=820.$ For each point in the diagram the
minimization was done with fixed $\kappa$ and $f.$ The three
different phases corresponding to the well known growth modes are
indicated in the diagram. See text for details. }

\end{figure}

We have calculated the global phase diagram (GPD) as shown in Fig.
\ref{global_phase_diagram} for $N_A=820$ and $l_w = 200$ (coverage
$\approx 4$ layers) in the $\kappa-f$ plane, where $\vert f\vert
\leq 7\%$ and $\vert\kappa\vert \leq 6\%$. Numerical data for this
diagram have been taken with steps of $0.5\%$ along both axes.

Following the approach of Scheffler {\it et al.} \cite
{Pehlke1997,Moll1998, Wang1999, Wang2000}, the growth process can be
divided into different stages. The early nucleation stage mainly
determines the island density. In our model, this is fixed by the
size of the unit cell $l_w$ which is inversely proportional to the
island density. The second stage is where the islands can grow at
the expense of the wetting layers. The main driving force here is a
pathway to lower the strain energy while not sacrificing too much
the adsorbate--substrate interface bonding energy.

The data in Fig. \ref{global_phase_diagram} are calculated by the
following steps. For a fixed $N_A$ at $820,$ we investigated all
configurations corresponding to an island with  shape corresponding
to A, B or C as shown in Fig. \ref{shapefig}, together with any
number of complete wetting film layers and at most one partial
wetting film layer (layers which are more than seven atoms wider
than the island base are counted as partially filled wetting
layers). The initial configuration has the equilibrium lattice
constant of the substrate. When the energy of the system is
minimized with molecular dynamics cooling, the adsorbate releases
strain energy by relaxation. This local minimization technique does
not lead to plastic deformations in this parameter range. Each point
of Fig. \ref{global_phase_diagram} corresponds to a relaxed
configuration with the lowest energy.

As can be seen in Fig. \ref{global_phase_diagram}, two well-known
phases already identified in the submonolayer regime are present in
different regions of the phase diagram for the higher coverage
situation: the Frank-van der Merwe (FM) phase dominates for
relatively small misfits and attractive substrate interactions while
the Volmer-Weber (VW) phase is favored for larger misfits and posive
values of $\kappa.$ As the lattice misfit $f$ increases, there is a
transition from wide, relatively thin islands to sharper forms.

However, perhaps the most interesting feature of the phase diagram
for coverages beyond two layers is the appearance of the
Stranski-Krastanov (SK) regime for large misfits and negative values
of $\kappa.$ In the SK regime, islands exist on top of wetting
films. With the present interaction parameters, the SK phase lies
between $\vert \kappa \vert \geq 1\%$ and $f \geq 4\%.$ The triple
points separating the phases are located approximately at
$(\kappa,f) = (0.5,4)\%$ and $(\kappa,f) = (-0.5,-3.5)\%$. Again,
the phase diagram is asymmetric between tensile and compressive
strain values, even for relatively small values of $f.$

We have numerically calculated the global phase diagrams for other
values of $N_A = 300$ and $1000$ and the results are very similar.
For coverage $N_A = 300$, the phase boundary between the FM and SK
regimes becomes less well defined. However, for $N_A = 1000,$ the
changes to Fig. \ref{global_phase_diagram} are minor; the positions
of the triple points seem to move to smaller values of $\vert f
\vert.$

\subsection{Optimal Shape and Size for Islands in the SK Mode}

\begin{figure}[t] \begin{center}
\centering\epsfig{file=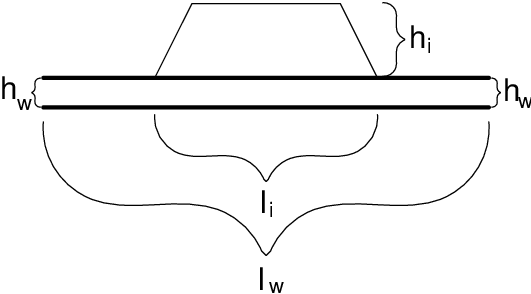,width=82mm}
\end{center}
\caption{A schematic figure of a typical island geometry considered
here. The height of the island and the wetting film are denoted by
$h_i$ and $h_w,$ respectively. The island--island separation, which
equals the width of the periodic simulation cell is denoted by
$l_w,$ and the island base width is $l_i.$ \label{shapepar}}
\end{figure}

For technological applications, it would be desirable to have a
self-organized array of islands which has a very narrow size and
shape distribution function. Experimentally, such ``optimal'' island
sizes have been observed in some heteroepitaxial systems
\cite{Jubert01, Fruchart06, Silly05, Hansen99}. The basic question
is whether the optimal size islands correspond to equilibrium
minimum energy configurations, or are they just a consequence of
kinetic limitations that prevent further growth on accessible time
scales. The existence of optimal shapes from purely energetic
considerations has been proven for some specific systems such as
those in Refs. \cite{Golovin04,Shchukin1995,Chiu04,Wang1999}. In
Fig. \ref{shapepar}, we define a number of parameters which
characterize the shape and size of the island and the wetting film.
The height of the island and the wetting film are denoted by $h_i$
and $h_w,$ respectively. The island--island separation, which equals
the width of the periodic simulation cell is denoted by $l_w,$ and
the island base width is $l_i.$ A schematic diagram is shown in Fig.
\ref{shapepar}.

\begin{figure}[t]
\centering\epsfig{file=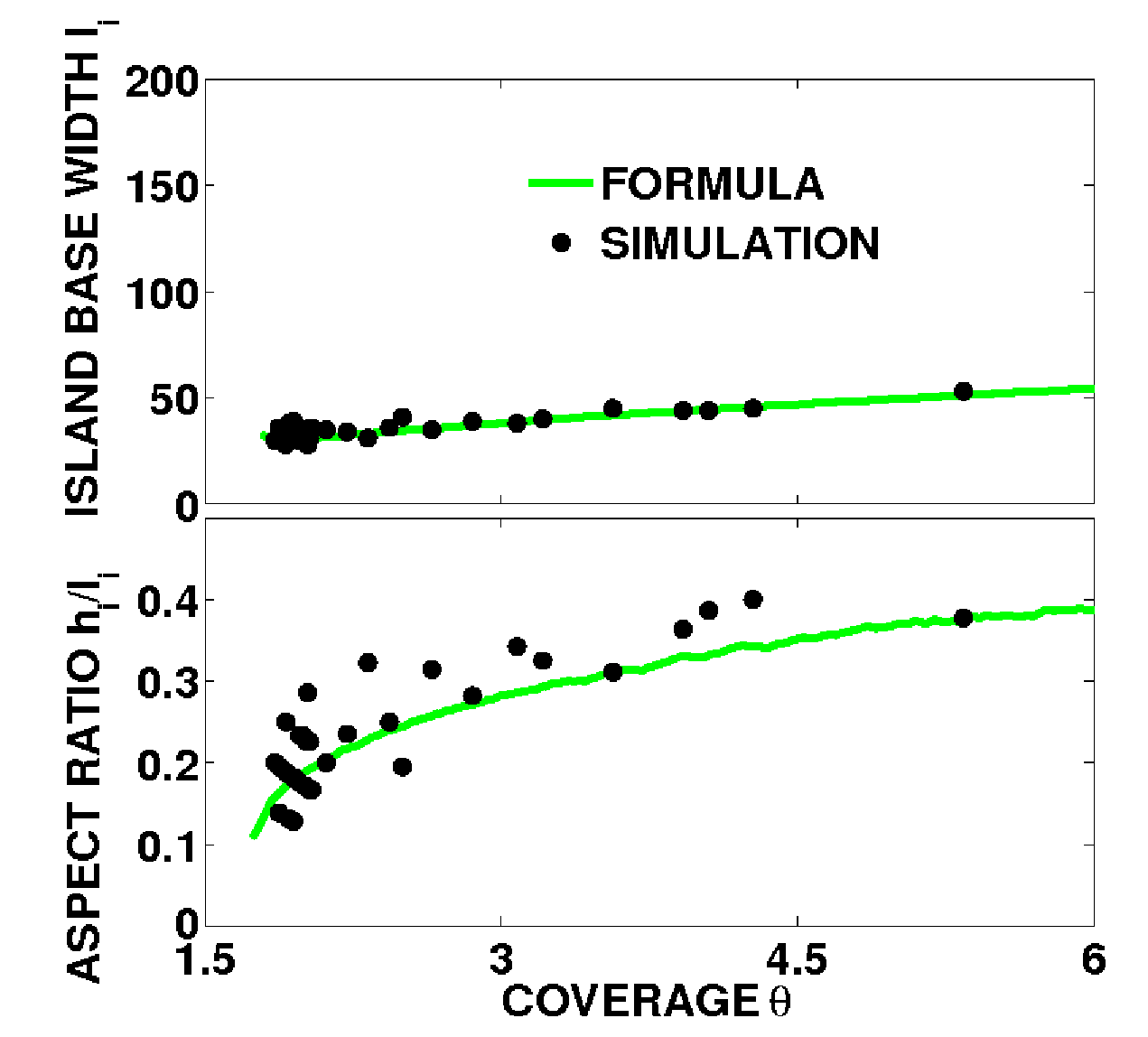,width=86mm}

\caption{\label{OPTIMAL_FIGURE}
Figure demonstrating the appearance of optimal island shapes and
sizes in the SK regime. The direct energy minimization data points,
which are marked with dots, are obtained by minimizing the energies
of all acceptable shapes of a given coverage and selecting the
lowest energy configuration among them. The solid line is an
estimate given by numerical minimization of the interpolation
formula.
}
\end{figure}

We have investigated the optimal shapes and sizes of the islands in
the SK mode for the misfit parameter $f = 5\%$ and interaction
parameter $\kappa = -4\%$ as a function of different amounts of
adsorbate $N_A$, for an island density corresponding to $l_w= 200$.
The results are shown in Fig. \ref{OPTIMAL_FIGURE}. At low coverages
$\theta = N_A/l_w$ the atoms completely wet the substrate and form
wide, non-wetting islands after which the actual SK regime emerges.
For coverage $\theta \ge 4$, both the base width $l_i$ and the
aspect ratio $h_{i}/l_i$ of the island increase slowly with
coverage. Typically, the adsorbate forms a combination of one full
and one incomplete wetting layer together with an island of shape A.

\begin{figure}[t]
\epsfig{file=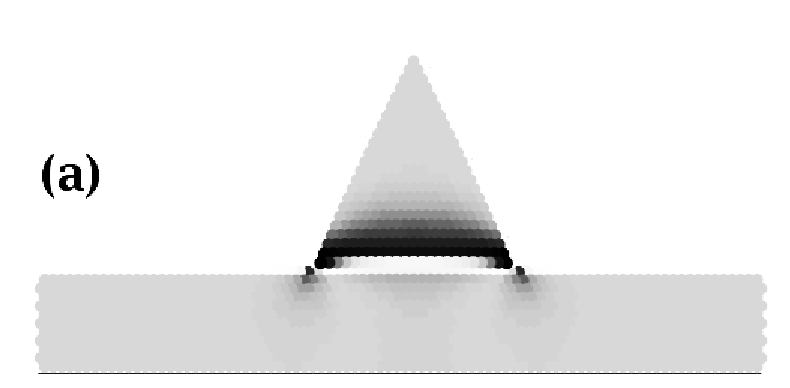,width=88mm}
\epsfig{file=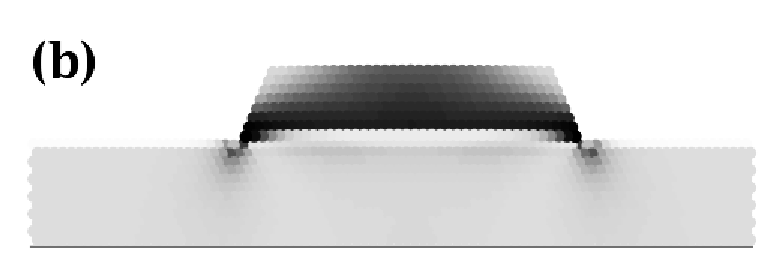,width=88mm}
\epsfig{file=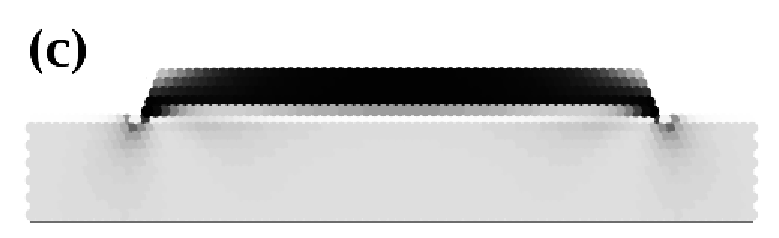,width=88mm}
\caption{\label{OPTIMAL_STRESS}
The qualitative change in the relaxation pattern in wide and narrow
islands. These shapes are taken from a larger configuration set
which was used to fit Eq. (\ref{omega}). The darker colors indicate
increasing levels of strain. To enhance contrast, the outmost
adsorbate-vacuum surface layer is not shown.  The configurations
shown here correspond to $N_A = 500.$ The islands have shape A on
top of one complete layer of wetting film. The widths $l_i$ are $24,
41$ and $62$ from top to bottom and the heights $h_i$ are $24, 8$
and $5,$ respectively. The configuration in the middle with aspect
ratio $h_{i}/ l_i=0.2$ and base width $l_i=41$ has the lowest
energy. }
\end{figure}

To understand the strain relaxation in the SK phase, we select the
$\theta = 2.5$ point from Fig. \ref{OPTIMAL_FIGURE} corresponding to
total number of adatoms $N_A=500$. The optimal shape and size of the
island in this case has $l_i = 41$ and $h_i = 8$ on top of one
complete layer of wetting film, see panel (b) of Fig.
\ref{OPTIMAL_STRESS}. In addition to this configuration, we show the
spatial energy distribution of the totally relaxed configurations
for two other geometries selected from the high and low aspect ratio
ends of the shape spectrum, namely configurations with $l_i = 24,
62$ and $h_i = 24, 5,$ respectively. The strain patterns of these
different configurations are shown in panels (a) and (c) of Fig.
\ref{OPTIMAL_STRESS}. We note from these three geometries that the
strain pattern falls into two different categories. For the large
aspect ratio ($l_i = 24,$ $h_i=24$), except for the immediate
adsorbate substrate interface region, the bulk of the island is
largely relaxed. In the other limit of a small aspect ratio ($l_i =
62,$ $h_i=5$), except for the surface and corner region, most of the
atoms inside the island are strained. For small islands, the
relaxation favors sharp islands but the adsorbate--substrate
interface bonding energy and the surface energy favor wide islands.
It turns out that the most efficient way of lowering the strain
energy while still maintaining a strong interface bonding energy is
to pick a geometry in between these two extremes. This balance
determines the aspect ratio and the optimal shape of the island.
When the coverage increases, the total strain energy can be lowered
by allowing the size of the island to increase at the expense of
reducing the thickness of the wetting film. The optimal size of the
island results from the balance between the strain relaxation of
larger island and the ``thinning'' energies of the wetting film.
Another factor that contributes to determining the optimal size of
the island is the indirect interaction between the islands mediated
by the substrate relaxation. These points are discussed in more
detail in next section.

\section{Analytic Interpolation Formula for Total Energy}

In this section, we will look in more detail into the various
competing mechanisms leading to the existence of an optimal size and
shape for the islands in the SK regime. We also develop an
approximate analytic interpolation formula that incorporates all
these mechanisms and allows a simple interpretation of our numerical
data. Our approach is in many respects similar to those in
Refs. \cite{uemura02,katsuno05,Combe2001,Thibault2004}.

\subsection{Reference Energy \label{reference_energy}}

To develop an analytic interpolation formula, we first choose a
reference system corresponding to a completely unstrained substrate
and completely strained adsorbate such that the in\--plane spacing
between atoms in each layer of the island is the same as that in the
substrate.  This reference system has three kinds of bonds:
unstrained, strained and sub\-stra\-te--ad\-sor\-ba\-te interface
bonds. The corresponding energies are $\epsilon_{\rm ub}$ for
unstrained bonds and $\epsilon_{\rm xa}$ and $\epsilon_{\rm ya}$ for
horizontal and vertical adsorbate bonds, respectively. The
sub\-stra\-te--ad\-sor\-ba\-te interface bonds, which are vertical,
are denoted by $\epsilon_{\rm yi}.$ These are obtained from Eq.
(\ref{LJpot}), and the details are given in the Appendix.

\begin{table}[b!p!]

\caption{ Reference energy expressions. The quantity $l_0 = l_w -
l_i$ when $h_w \geq 1$ and $l_0=l_f$ when $h_w<1$ and $l_f$ is the
width of the partial wetting fraction, which is not under the
island. The symbols $\epsilon_{\rm ub}, \epsilon_{\rm xa}$ and
$\epsilon_{\rm ya}$ refer to unstrained substrate bond and strained
horizontal and vertical adsorbate bond energies, respectively.  The
symbol $\epsilon_{\rm yi}$ is the energy of a strained
substrate--adsorbate bond.}

\begin{tabular}{|l|c|}
\hline
 Contribution    &  Expression \\
  \hline \hline
{\bf Substrate bulk}    $E_{B}$    & $-6\epsilon_{\rm ub}N_S$  \\
  \hline \hline
{\bf Adsorbate bulk}  $E_{W}$    & $-(2\epsilon_{\rm xa} + 4\epsilon_{\rm ya})N_A$  \\
  \hline \hline
{\bf Surfaces} $E_{S}$ :   &  \\
  \hline
  \hline
 Substrate-vacuum   & $2\epsilon_{\rm ub}(l_w - l_i - l_0)$  \\
  \hline
 Wetting film-vacuum   & $2\epsilon_{\rm ya}l_0$  \\
  \hline
 Oblique facet-vacuum   & $2(\epsilon_{\rm ya}+\epsilon_{\rm xa})h_i$ \\
  \hline
 Island top-vacuum   & $2\epsilon_{\rm ya}(l_i - h_i)$ \\
  \hline
 Substrate-adsorbate  & $( 2\epsilon_{\rm ub} + 2\epsilon_{\rm ya}-4\epsilon_{\rm yi})(l_0+l_i)$ \\
  \hline
\end{tabular}
\label{reference_table}
\end{table}

The bulk contributions $E_B$ and $E_W$ for the substrate and
adsorbate are proportional to the number of atoms in the adsorbate
and substrate, respectively. In addition, each surface and interface
leads to a correction term due to missing or changed bond energies,
resulting in a total surface energy $E_S.$ For the wetting film,
partial layers are allowed to account for the situation where the
number of atoms in the wetting film deviates from multiples of
$l_w.$ All corner energies are neglected. The different
contributions to $E_B, E_W,$ and $E_S$ in terms of the bond energies
$\epsilon_{\rm ub}, \epsilon_{\rm xa}, \epsilon_{\rm ya},$ and
$\epsilon_{\rm yi}$ are listed in Table \ref{reference_table} above.

\subsubsection{Relaxation Energy of Island}

The total energy of a fully strain relaxed island can now be written
as $E_{\rm total} = E_{\rm ref} + \Omega$, where $\Omega$ is the
relaxation energy. As the panel (c) of Fig. (\ref{OPTIMAL_STRESS})
shows, the relaxation is seen only in the vicinity of the non-horizontal
facets when the island aspect ratio is low. In this situation the
top facet and the interior are both completely stressed. When the
qualitative pattern of relaxation is of this kind, the energy
difference $\Omega$ has to become practically independent of the
island width for a set of islands with the same height. If we
consider the islands with constant height in the high aspect ratio
case, the panel (a) of Fig. (\ref{OPTIMAL_STRESS}) shows that the
relaxation energy should depend almost linearly on the width.

To quantitatively verify this, we study a set of islands with
different values of island base widths $l_i$ and island heights
$h_i.$ The wetting layer thickness $h_w$ and the ratio $l_w/l_i$ are
kept constant with $h_w = 2$ and $l_w/l_i = 10$. This value of
$l_w/l_i$ is large enough so that the interaction energy between
islands is negligible at this separation. The chosen island heights
$h_i$ are $5, 15, 31, 45$ and $59.$ The base widths $l_i$ for each
height $h_i$ are chosen with a denser mesh near the large aspect
ratio limit and with a sparser mesh in the small aspect ratio,
shallow island limit. The energy of the island in each case is
minimized and then the additional relaxation energy is obtained
after the subtraction of the reference energy. The numerical data of
the extra relaxation energy for the entire set of islands with
different aspect ratios can be fitted to a simple analytic form
which has the properties described above, namely
\beq \label{omega} \Omega =E_{\rm shallow} - b(l_i - l_c -
\sqrt{((l_i - l_c)^2 + d^2}). \eeq
The first term $E_{\rm shallow}$ in Eq.(5) represents the relaxation
energy in the shallow island (small aspect ratio) limit, which can
be fitted to a form independent of $l_i$ given by
\beq
 \label{shallow}
 E_{\rm shallow} = -0.194(7)h_i(h_i + 8.16(7)) \ \epsilon_{\rm ss}.
\eeq
Here the term in $h_i^2 $ is related to the volume relaxation and
the term linear in $h_i$ comes from the surface relaxation of the
shallow island. It should be noted that when the island base $l_i$
is in the regime where $\Omega$ depends only weakly on $l_i$ and the
height $h_i$ becomes large, Eq. (\ref{shallow}) breaks down. In this
large volume limit the term is linearly dependent on $h_i,$ {\it
i.e.} it has turned into a correction to the surface energy of the
vertical island facets.

The other variables  $l_c$, $b$, and $d$ of Eq. (\ref{omega}) are
functions of the height of the island and the numerical fit of these
variables are given in Table \ref{numerical_table}. The value of
$l_c$ defines a crossover between the shallow island limit and the
sharp island limit where the qualitative nature of the relaxation
pattern in the island changes. For a fixed $h_i$, the energy of the
island starts to increase sharply when $l_i$ is reduced below the
threshold value $l_c$ which is a function of $h_i$.

To keep the analysis simple, we have not included small
contributions to the total energy, such as corner energies and
stretching energy of the adsorbate--substrate interface in the
interpolation formula. There exist rare circumstances when these
effects can become important and in these cases the formula does not
hold.

It should be noted that at some coverage the island volume becomes
necessarily so large that the relaxed neighborhoods of the left and right
facets do not overlap even in the equilateral case. In this limit
the relaxation energy is proportional to the length of the non-horizontal
facets and can be adsorbed to the corresponding surface tensions. In
this case the above formula is not expected to hold.

A similar study concerning the stability of strained heteroepitaxial
systems has been conducted in Ref. \cite{Thibault2004}. They find
that an island array is unstable in all the cases considered.
The main difference to our study is that we do not assume
the island-island distance to be fixed by the interaction between the
islands but by the nucleation stage, as in Refs.
\cite{Moll1998,Wang1999,Wang2000}.

\begin{table}[t!p!]
\caption{Parameters of Eq. (\ref{omega}) as function of height.}
\begin{tabular}{|c|c|}
\hline
 Parameter       &  Expression \\
  \hline \hline
  $  b              $   & $ 0.0134(2)(h_i + 1.602(2)) $ \\
  $l_c              $   & $ 2.45(7)h_i + 8.47(3)      $ \\
  $d                $   & $ 1.21(0)h_i + 8.82(0)      $ \\
  \hline
\end{tabular}
\label{numerical_table}
\end{table}

\subsubsection{Comparison of Numerical Data with Interpolation Formula}

\begin{figure}[t] \begin{center}
\centering\epsfig{file=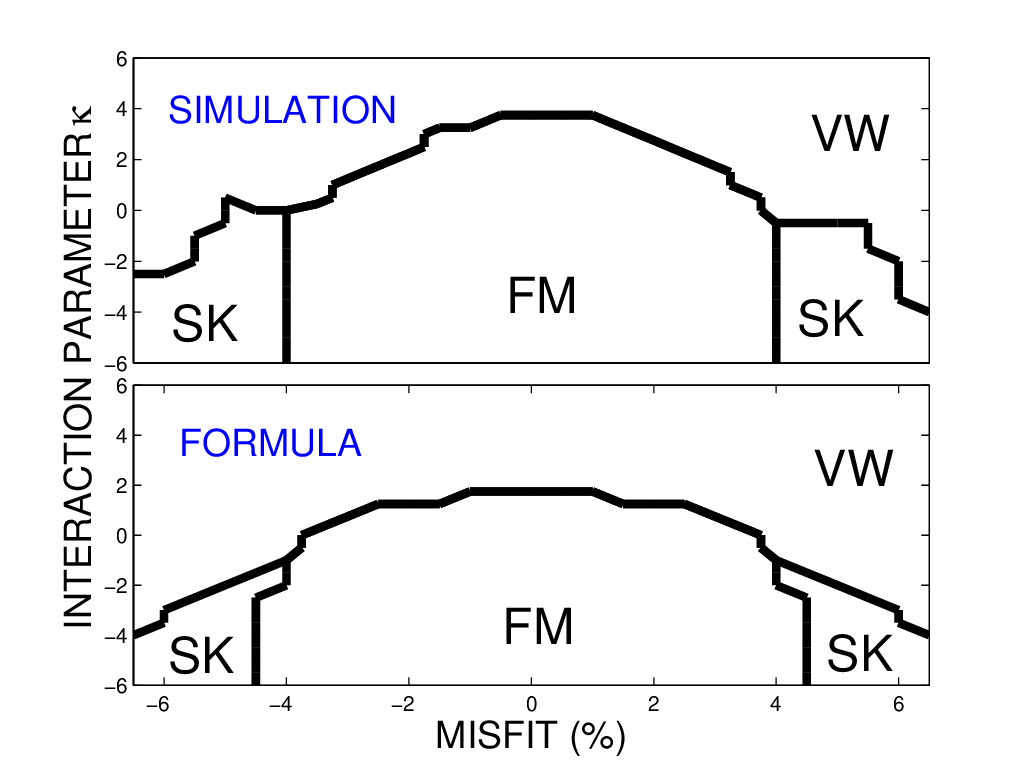,width=92mm}
 \end{center}
\caption{\label{compared_global_grams}
The analytic interpolation formula for the total energy yields a
global phase diagram in good agreement with one obtained from direct
numerical minimization. The fitting of the formula was done in the
point where $f = 5\%$ and $\kappa = -4\%.$ The small asymmetry of
the direct minimization is related to the asymmetry of the
interaction potential which is especially pronounced near the
locations of the largest displacements such as the corners, as Fig.
\ref{OPTIMAL_STRESS} demonstrates. In the large scale the direct
minimization result is fairly symmetric, which suggests that the
rest of the relaxation energy scales roughly as $f^2$ as the
continuum theory predicts. Since the interpolation formula already
neglects the corner energies, we also take the elastic energy
scaling to be of the $f^2$ form. This explains the mirror symmetry
of the positive and negative misfit sides of the interpolated phase
diagram in the lower panel.  }
\end{figure}

In this section, we will use the interpolation formula
obtained in the previous section to examine the global phase diagram
and determine the optimal sizes and shapes of the islands in the SK
regime. First, we examine the global phase diagram. For this
purpose, we fix the total number of adsorbate atoms $N_A = 820$ and
determine the minimum energy configuration as a function of the
misfit parameter $f$ and interaction parameter $\kappa$
by comparing the total energies of different configurations. Note
that the reference energy $E_{\rm ref}$ already has a dependence on
$\kappa$ and $f.$ For the additional relaxation energy $\Omega$ we
take the formula, which was
fitted for $f = 5\%,$ $\kappa = -4\%.$ We verified numerically that
in the first approximation it scales with the misfit as $f^2$ and
neglect its dependence on $\kappa$ \cite{Footnote1}. The resulting
GPD as shown in Fig. \ref{compared_global_grams} is very close to
the one obtained from the direct energy minimization as shown in Fig
\ref{global_phase_diagram}. The disagreements have generally three
sources.

First, the interpolation formula is constrained to islands of shape
A in some points of the phase diagram. This can lead to two kinds of
differences. As we showed earlier, when all configurations are
compared with the direct minimization, the lowest energy shape is
almost always among A, B, or C shapes so that the extra facets in
the B and C shapes are of the shortest possible length. However, at
some coverages the combinatorics limits the number of these three
shapes to be fairly small. It can happen that these shapes are also
of extremal widths or thicknesses whose leading energy contributions
are out of the average range. Under these circumstances the
configuration space would be sampled better by allowing the shapes
to have some asymmetry or additional facets at the corners. This
finite size effect is relevant for small island volumes only, and as
we showed it is not very significant even in this regime. Anyhow, if
the interpolation formula is minimized analytically, the
configuration space is continuous and does not suffer from this
finite size effect. Another kind of difference originating from the
same source is the fact that sometimes the small terms (such as the
corner energy) not present in the formula can force the direct
energy minimization to switch to a wider island width at a lower
coverage than what the interpolation formula gives. This can reduce
the island height by one and increase the width by two layers but
the effect is not cumulative and cannot separate the direct
minimization and interpolation results far apart.

Second, close to the critical coverage separating two growth modes
the energies of very different configurations, such as a flat layer
and an island, are comparable. In this kind of situation the small
energy terms not present in the interpolation formula can actually
determine the energy balance and the real lowest energy shape is
therefore not detected by the formula. However, when the coverage is
outside the transition lines these cases become irrelevant.

Third, the analytic formula is less reliable in the large aspect
ratio limit (sharp islands) with high coverages. The height
dependence of the parameters is fitted for the small island limit
where the surface and bulk energetics are not clearly separable.
With sufficiently large island volume the relaxation becomes clearly
associated with the island facets and in this case the relaxation
energy must be directly proportional to the length of the associated
facets. Since in this case the stress energy cost increases
proportionally to the island volume, dislocations nucleation becomes
relevant; an effect which is not included in the present model. In
this limit also the validity of the $f^2$ scaling of the relaxation
energy breaks down.

Finally, we should also mention that the island--island interaction
energy is neglected in the present interpolation formula.  However,
we have checked that this effect is negligible at a unit cell length
of $l_w=200$ and does not affect the results.

Next, we investigate the optimal size and shape of the islands in
the SK regime for the parameters $f = 4\%$ and $\kappa = -3\%.$ For
this purpose, we take the interpolation formula for $\Omega$ derived
for a fixed thickness for the wetting film $(h_w=2)$ and apply it to
configurations for other values of wetting film thickness. This
amounts to neglecting the wetting film thinning energy and is only
expected to be valid in the low coverage limit. We vary the size and
aspect ratio of the island for a given coverage through choosing
different values for $l_i$ and $h_i.$ The corresponding value for
the thickness of the wetting film $h_w$ is then determined by the
constraint of the total amount of the adsorbate material being
constant. The total energy expression
is minimized 
to yield the optimal value of $l_i$ and $h_i$, which determine the
optimal size and shape for a given coverage.

In Fig. \ref{OPTIMAL_FIGURE} we show the island base width $l_i$ and
aspect ratio $h_{i}/l_i$ as a function of the coverage. The
predicted optimal island base widths from the direct numerical
calculation and those obtained from the total energy minimization
with the interpolation formula are close. 
The optimal aspect ratios determined from the direct minimization and
from the interpolation formula have the same qualitative coverage
dependence.

\section{Summary and Conclusions}

In this work, we have investigated the various phases resulting from
the relaxation of strain energy starting from initial epitaxial
strained adsorbate wetting films. We employ the molecular static
method which is a powerful tool to overcome the restrictions related
to conventional minimization methods that rely on the continuity and
differentiability of the underlying manifold. In this work the phase
diagram of 2D heteroepitaxial thin film systems was obtained by
direct minimization for various combinations of the lattice misfit
$f$ and parameter $\kappa$ describing the interface tension. In the
submonolayer regime, the Stranski-Krastanow (SK) phase is not
present, but a transition between layer-by-layer (FM) equilibrium
shapes and Volmer-Weber (VW) shapes is sensitive to the value of
$\kappa.$ As the adsorbate coverage increases, the SK phase appears
between the FM and VW regions when $\kappa$ had a value that was
favorable for the contact between substrate and adsorbate. The
islands always have truncated pyramid shapes with minor edge
defects.

In the SK regime, we have investigated the important issue
concerning the optimal shape and size of islands by searching
through the minimum energy configurations of systems for fixed
material parameters $f$ and $\kappa.$ The shapes were restricted to
truncated pyramids. The actual optimal size and aspect ratio depend
on the coverage of the adsorbate atoms. Both the aspect ratio and
the base width of the island increases slowly with coverage. To
understand better the numerical results, an analytic interpolation
formula was developed. This approximate formula takes into account
the strain relaxation from a reference state of totally strained
wetting films, adsorbate island and an unstrained substrate. It
produces results that agree very well with the numerical energy
minimization data.

The optimal shape and the aspect ratio result from a compromise of
increasing the relaxation energy for a sharp (higher aspect ratio)
island {\it vs.} increasing the adsorbate--substrate interaction for
a shallow (low aspect ratio) island. The optimal size of the island
grows as the coverage increases because a larger island is more
effective in strain relaxation and lowering of the elastic energy.
Eventually, when the base width of the island $l_i$ grows to be
comparable with the unit cell size $l_w$, the optimal island size
will be determined by the substrate mediated indirect interaction
between the islands, as suggested in Ref. \cite{Shchukin1995}. It is
important to note that the details of the coverage dependence of the
optimal size in our model differ somewhat from the results of Ref.
\cite{Wang2000}, where the wetting film thinning energy was
evaluated using a continuum approach and is finite at thickness even
above two layers. In our model, the wetting film thinning energy
implicitly present in both the interpolation formula and the
numerical energy minimization is negligible until the last layer of
wetting film starts to get depleted. Thus the optimal size at low
coverages shown in Fig. \ref{OPTIMAL_FIGURE} corresponds to the
total adsorbate atoms minus one complete plus possible one partial
wetting film layer.

Finally, we would like to discuss the applicability of our
results for realistic 3D cases. While the fact that our global phase
diagram correctly reproduces all the well-known growth modes in the
relevant regimes indicates the overall validity of the model, there
are several additional features in 3D, which need to be considered.
Assuming no intermixing of the substrate and film atoms, the most
important ingredient missing in 2D is the spatial anisotropy of
various crystal surfaces. Recent experiments in 3D indicate
\cite{Silly05} that changing the mismatch of the substrate can
induce significant morphological changes in 3D islands. For the
close-packed (111) surface geometry, which most closely corresponds
to the 2D case studied here, we have done minimization of selected
island shapes and find that there are stable island configurations,
which are straightforward 3D generalizations of the 2D shapes found
here. However, a systematic search for optimal 3D nanoisland shapes
is computationally very expensive and beyond the scope of the
present work.

\section{Acknowlegdements:}
This work acknowledges joint funding under EU STRP 016447 MagDot and
NSF DMR Award No. 0502737. J.J. and T.A-N. acknowledge support from
the Academy of Finland through its Center of Excellence COMP grant
and from Finnish Center of Scientific Computing (CSC) through computing
resources. J.J. also acknowledges support from the Foundation of Vilho,
Yrj\"{o} ja Kalle V\"{a}is\"{a}l\"{a}n rahasto of Finnish Academy of
Science and Letters. E.G. acknowledges support from FAPESP under
Grant No. 07/08492-9.

\section{Appendix: Bond Energies In Epitaxial Strained Film}

An adsorbate bulk atom has two horizontal and four diagonal bonds,
the energies of which are denoted by $\epsilon_{\rm xa}$ and
$\epsilon_{\rm ya},$ respectively. The horizontal distance of the
atoms in a coherent adsorbate layer is $r_{\rm ss}$ and the
equilibrium distance is $x'_0 = r_{\rm ss}(1 + f).$ The energy of a
horizontal bond $-\epsilon_{\rm xa}$ in the reference state is thus
$V_{\rm ff}(r_{\rm ss})/2.$ In a hexagonal lattice with pair
potential the Poisson ratio is always $\nu = 1/3.$ Thus, the ratio
of vertical and horizontal strains $u_{yy}$ and $u_{xx}$ is
\beq \frac{u_{yy}}{u_{xx}} = -\frac{1}{3} \eeq
or
\beq u_{yy} = \frac{y'}{y'_0} - 1 = \frac{x'_0-r_{\rm ss}}{3x'_0}
\eeq
where $y'$ is the actual distance between adjacent horizontal layers
in a strained adsorbate and $y'_0$ is the corresponding equilibrium
distance, $y'_0 = \sqrt{3}x'_0/2.$ This gives
\begin{align}
y'  =  y'_0\left(\frac{4x'_0 - r_{\rm ss}}{3x'_0}\right)
\end{align}
The length of the diagonal bond $r_d$ in the initial strained film
is given by the relation $r_d^2 = r_{\rm ss}^2/4 + 3(y')^2/4$ and
energy of the bond $-\epsilon_{\rm ya}$ is given by $V_{\rm ff}/2$
of Eq. (\ref{LJpot}) with the argument $r_d.$

The bond strength between the substrate and adsorbate $\epsilon_{\rm
yi}$ can be easily obtained from the adsorbate bond $\epsilon_{\rm
ya}$ with two modifications. First, the misfit parameter $f$ is
replaced by $f/2$ and then the overall strength is multiplied by a
factor $1 + \kappa.$ The first modification arises from the fact
that the equilibrium distance between the substrate and adsorbate
atom is the arithmetic mean of the equilibrium distances between
pure materials and the second follows from the definition of
$\kappa$ relating the strength of the film--substrate interaction to
the film--film interaction.


\begin{thebibliography}{15}
\bibitem{Footnote1} By calculating $\Omega$ for several values of $l_i, h_i$  and $f$ we saw that 
the misfit-dependence of $\Omega$ is essentially a shape-independent prefactor.
Furthermore, we verified that other terms than $f^2$ can be safely
omitted. It is possible that this kind of misfit scaling is a feature of 
the present model.
We also varied $\kappa$ and found that the dependence of $\Omega$ on $\kappa$ is 
orders of magnitude too small to have any relevance here.
\end{thebibliography}
\end{document}